\documentclass[%
preprint,
 amsmath,amssymb,
 aps,
 ]{revtex4-1}
 \usepackage{graphicx}
\usepackage{amsmath}
\usepackage{slashed}
\usepackage{amsfonts}
\usepackage{amsmath}
\usepackage{booktabs}
\usepackage{siunitx}
\usepackage{multirow}
\usepackage{colortbl}
\usepackage{xcolor}
\usepackage[%
  colorlinks=true,
  urlcolor=blue,
  linkcolor=blue,
  citecolor=blue]{hyperref}

\begin{document}
\raggedbottom
\title{Light-Cone Distribution Amplitudes of Light $J^{PC}=2^{--}$ Tensor Mesons in QCD% \tnoteref{mytitlenote}
}

\author{T.M.Aliev}
  \email{taliev@metu.edu.tr}
 \author{S. Bilmis}%
   \email{sbilmis@metu.edu.tr}
\affiliation{%
Department of Physics, Middle East Technical University, 06800, Ankara, Turkey
}%

\author{Kwei-Chou Yang}
\email{kcyang@cycu.edu.tw}
\affiliation{Department of Physics and Center for High Energy Physics, Chung Yuan Christian University, Taoyuan 320, Taiwan}

\date{\today}% It is always \today, today,

\begin{abstract}
We present a study for two-quark light-cone distribution amplitudes for the $1^3D_2$ light tensor meson states with quantum number $J^{PC}=2^{--}$. Because of the G-parity, the chiral-even two-quark light-cone distribution amplitudes of this tensor meson are antisymmetric under the interchange of momentum fractions of the quark and antiquark in the SU(3) limit,  while the chiral-odd ones are symmetric.  
The asymptotic leading-twist  LCDAs with the strange quark mass correction are shown. We estimate the relevant parameters, the decay constants $f_T$ and $f_T^\perp$, and first Gegenbauer moment $a_1^\perp$, by using the QCD sum rule method. 
These parameters play a central role in the investigation of $B$ meson decaying into the $2^{--}$ tensor mesons.
\end{abstract}

% \pacs{}% PACS, the Physics and Astronomy
\maketitle

\section{Introduction}
\label{sec:intro}
Analysis of spectroscopy of particles represents promising area provides to check predictions of quantum chromodynamics in perturbative and nonperturbative domains.
During recent years inspiring results are obtained in this area; namely, new charmonium states are observed in experiments~\cite{Zhu:2007wz,Swanson:2006st,Rosner:2006sv}. The main result of these experimental studies is that the structures of these new states are not described by the conventional quark-antiquark picture, and they have more complex structures. About the studying properties of these states, the molecular picture, the tetraquark picture, the hybrid charmonium, or the baryonium states are used.

In the investigation of the spectroscopy of mesons with $J=0;1$ in QCD sum rules method the local interpolating currents without derivative $\bar{\Psi} \Gamma \Psi$ are used, where $\Gamma=1,~\gamma_5~,\gamma_\mu,~\gamma_\mu~,\gamma_5$. For studying the higher spin states the current with derivative is necessary. The interpolating current for light unflavored spin \rm{2}-mesons first time was constructed in~\cite{Aliev:1981ju}, and the results are relevant to the Gegenbauer moments in determination of light-cone distribution amplitudes of these states ~\cite{Cheng:2010hn}.  Some properties of the tensor mesons have been studied. For instance, the mass and decay constant of strange tensor meson $K_2^*(1430)$ by taking into account $SU(3)$ symmetry breaking effects within QCD sum rules is studied in~\cite{Aliev:2009nn}, and the mass and decay constant of heavy $\chi_{Q_2}$ tensor mesons within the same framework is determined in~\cite{Aliev:2010ac}.  

 The mass spectra of the negative parity tensor mesons $2^{--}$, containing light-light, light-heavy and heavy-heavy quarks, were calculated in~\cite{Chen:2011qu} in the QCD sum rule approach. 
The conventional quark model has predicted the existence of the $1^3D_2$ states with quantum number $J^{PC}=2^{--}$.
However, the light $2^{--}$ meson states, except $K_2(1820)$ meson, have not been observed yet.  Theoretically, it is thus important to measure the full set of this quantum states.  In this work, we will focus on the study of the light-cone distribution amplitudes (LCDAs) of the light $2^{--}$ meson states. Its LCDAs are relevant to the search of this particle from the $B$ decays in $B$-factories. In the past few years, $B$ decays involving  a light $2^{++}$ tensor meson had been observed \cite{delAmoSanchez:2010qa,Aubert:2006fj,Aubert:2009sx,Aubert:2008bj,Garmash:2005rv,Aubert:2005ce,Garmash:2004wa,Aubert:2008bc,Aubert:2009av,Aubert:2007bs,Aubert:2009me,Garmash:2006fh,Aubert:2008zza}. It will be very interesting to observe the light $2^{--}$ meson state in the $B$ decay from the theoretically point of view due not only to the prediction in the QCD-based quark model but also to clarification of rate deficit and polarization puzzles \cite{Cheng:2010yd}.

The properties of LCDAs for $2^{--}$ states are quite different from the corresponding ones for $2^{++}$ and $2^{-+}$ states. For instance, in SU(3) limit, under interchange of two quarks' momentum fractions, the leading LCDA $\phi_\parallel$ is anti-symmetric, and $\phi_\perp$ is symmetric for $2^{--}$ states. Nevertheless, correspondingly, for $2^{-+}$ states,  the former is symmetric and the latter is anti-symmetric, while for  $2^{++}$ states, these two leading LCDAs are anti-symmetric.  Once the leading LCDAs are obtained, the twist-3 two-quark LCDAs can be computed by means of Wandzura-Wilczek relation.

As what were done in~\cite{Chen:2011qu}, for simplicity, we will neglect the possible mixtures between two strange states, $K_2 (1820)$ (the $^1D_2$ state) and $K_2(1770)$ (the $^3 D_2$ state), and between ${\bar s}s $ and ${\bar u} u + {\bar d} d$, where the former is due to the mass difference of strange and light quarks.  If in the near future we can find out such states with mass $\sim1.9$~GeV, the mixing angle can be estimated by means of Gell-Mann Okubo relation.  (The result is similar to the case of $K_1(1270)$ and $K_1(1400)$, and that of $f_1(1285)$ and $f_1(1420)$ \cite{Yang:2007zt,Cheng:2013cwa}.)  A further precise estimate will be then important.

The rest of the paper is organized as follows. In section~\ref{sec:lightcone}, we define light-cone distribution amplitudes for the $2^{--}$ meson states and discuss their properties. The (Wandzura-Wilczek) relations between twist-3 LCDAs and leading-twist LCDAs are given. We will approximately parametrize the leading leading LCDAs in terms of Gegenbauer polynomials up to the term containing the first Gegenbauer moment, where the relevant parameters, $f_T, f_T^\perp$ and $a_1^\perp$, defined in section~\ref{sec:lightcone}, are estimated by means of the QCD sum rule approach, and the results are given in section~\ref{sec:calculation}.  
Section~\ref{sec:numerical} is devoted to the numerical calculations of these parameters. To estimate the possible uncertainty in the obtained QCD sum rules due to the finite width of the $2^{--}$ states, we will include $\pm 150$~MeV uncertainty to the excited state threshold $\sqrt{s_0}$ that models the contribution from higher states. Such uncertainty range is compatible with  the widths of $K_2(1820)$ and other higher resonance states. We summarize in section~\ref{sec:summary}.

\section{Light-cone distribution amplitudes}
\label{sec:lightcone}
We define the chiral-even light-come distribution amplitudes of a light tensor meson with quantum number $J^{PC}=2^{--}$  to be 
\begin{eqnarray}
\label{eq:chial-even-LCDAs}
&& \langle T(P,\lambda)|\bar q_1(z)\gamma_\mu\gamma_5  {q_2}(-z)|0\rangle
 =   i f_{T} m_T^3
  \int\limits_{0}^1 \! du\,e^{i(u-\bar u)pz} \Bigg\{ p_\mu \frac{\epsilon^{(\lambda)*}_{\alpha\beta}z^\alpha z^\beta}
{(pz)^2} \,\phi_\parallel(u)
\nonumber \\
&&  + \frac{\epsilon^{(\lambda)*}_{\perp\mu\alpha} z^\alpha }{pz}\, g_a(u)
 - \frac{1}{2} z_\mu \frac{\epsilon^{(\lambda)*}_{\alpha\beta}z^\alpha z^\beta}{(pz)^3} m_T^2 \,g_3(u)
\Bigg\}\,,
\\
&& \!\!\!\! \!\!
 \langle T(P,\lambda)|\bar q_1(z)\gamma_\mu q_2 (-z)|0\rangle
   =  i f_{T} m_T^3
  \int\limits_{0}^1 \! du\,e^{i(u-\bar u)pz} \varepsilon_{\mu\nu\alpha\beta} \frac{z^\nu p^\alpha}{pz}
   \epsilon_{(\lambda)}^{*\beta\delta}  z_\delta\,  \, g_v(u)\,, \nonumber\\
\end{eqnarray}
and its chiral-odd LCDAs to be
\begin{eqnarray} 
\label{eq:chial-odd-LCDAs}
&& \langle T(P,\lambda)|\bar q_1(z)\sigma_{\mu\nu} \gamma_5 q_2(-z)|0\rangle
  =
  f_{T}^\perp m_T^2
  \int\limits_{0}^1 \! du\,e^{i(u-\bar u)pz} \Bigg\{  
\nonumber\\
&&  \ \ \ 
   \left[\epsilon^{(\lambda)*}_{\perp\mu\alpha} z^\alpha p_\nu
 - \epsilon^{(\lambda)*}_{\perp\nu\alpha} z^\alpha p_\mu\right] \frac{1}{pz} \phi_\perp(u)  
  +  (p_\mu z_\nu - p_\nu z_\mu)
 \frac{ m_T^2\epsilon^{(\lambda)*}_{\alpha\beta}z^\alpha z^\beta}{(pz)^3} h_t(u)   
 \nonumber\\
 && \ \ \ 
  +   \frac{1}{2} \left[\epsilon^{(\lambda)*}_{\perp\mu\alpha} z^\alpha z_\nu
 - \epsilon^{(\lambda)*}_{\perp\nu\alpha} z^\alpha z_\mu\right] \frac{m_T^2}{(pz)^2} h_3(u)
  \Bigg\} \,,
  \\
 && \langle T(P,\lambda)|\bar q_1(z) \gamma_5 q_2(-z)|0\rangle
 =
f_{T}^\perp m_T^4
  \int\limits_{0}^1 \! du\,e^{i(u-\bar u)pz} \frac{\epsilon^{(\lambda)*}_{\alpha\beta}z^\alpha z^\beta}{pz}h_p(u)\,, 
\end{eqnarray}
where $u$ and $\bar u\equiv 1-u$ are the momentum fractions carried by $q_1$
and $\bar q_2$ quarks,  respectively,  in the meson.
$\phi_\parallel, \phi_\perp$ are leading twist-2 LCDAs, 
$g_v, g_a, h_t, h_p$ are twist-3 ones, and $g_3$ and $h_3$ are of twist-4. 
Here $z_\mu$ and $p_\nu\equiv P_\nu - z_\nu m_T^2/(2pz)$ are the two light-like vectors, with $P_\nu$ and $m_T$ being the momentum and the mass of the tensor meson, respectively. Using these two light-light vectors, the transverse component of the tensor meson's polarization $\epsilon^{(\lambda)}_{\mu\nu} $ can be given by
\begin{eqnarray}\label{eq:polprojectiors}
 && \epsilon^{(\lambda)}_{\perp\, \mu\nu} z^\nu
        = \epsilon^{(\lambda)}_{\mu\nu} z^\nu -
     \frac{\epsilon^{(\lambda)}_{\alpha\nu} z^\alpha z^\nu }{p z} \left( p_\mu-\frac{m_T^2}{2 p z} \,z_\mu\right)\,.
\end{eqnarray}

Due to the $G$-parity, in SU(3) limit,  $\phi_\parallel, g_a, g_v$, and $g_3$ are antisymmetric under the replacement $u\to 1-u$, while $\phi_\perp, h_t, h_p$ and $h_3$ are symmetric.  We do not further consider $g_3$ and  $h_3$ here. Neglecting the three-parton distribution amplitudes containing gluons, twist-3 LCDAs $g_a,g_v,h_t$, and $h_p$ are related to twist-2 ones through the Wandzura-Wilczek relations:
\begin{eqnarray}\label{eq:WW}
    g_a^{WW}(u) &=& \int\limits_{0}^u dv\, \frac{\widetilde{\phi}_\parallel(v)}{\bar v}+
                  \int\limits_{u}^1 dv\, \frac{\widetilde{\phi}_\parallel(v)}{v}
                  +  \tilde\delta_- \phi_\perp\,,
\nonumber\\
    g_v^{WW}(u) &=& 2\bar{u}\int\limits_{0}^u dv\, \frac{\widetilde{\phi}_\parallel(v)}{\bar v}+
                  2u\int\limits_{u}^1 dv\, \frac{\widetilde{\phi}_\parallel(v)}{v}\,,
\nonumber\\
 h_t^{WW}(u) &=& \frac{3}{2} (2u-1)\left(\int\limits_{0}^u dv\, \frac{\widetilde{\phi}_\perp(v)}{\bar v} -
                  \int\limits_{u}^1 dv\, \frac{\widetilde{\phi}_\perp(v)}{v}\right)   +  \delta_- \phi_\parallel\,,
                    \nonumber\\
    h_p^{WW}(u) &=& 3 \left( \bar{u}\int\limits_{0}^u dv\, \frac{\widetilde{\phi}_\perp(v)}{\bar v}+
                  u\int\limits_{u}^1 dv\, \frac{\widetilde{\phi}_\perp(v)}{v} \right)\,,
\end{eqnarray}
where 
\begin{eqnarray}
\widetilde{\phi}_\parallel (v) & =& \phi_\parallel (v) +\frac{1}{4} \tilde\delta_- \xi \phi_\perp^\prime (v)  
                                                     -\frac{1}{4} \tilde\delta_+  \phi_\perp^\prime (v)\,, \nonumber\\
\widetilde{\phi}_\perp (v) & = &  \phi_\perp (v) - \frac{1}{3} \delta_- \left(\phi_\parallel (v) - \frac{1}{2} \xi \phi_\perp^\prime (v)  \right)
           -\frac{1}{2} \delta_+  \phi_\parallel^\prime (v)\,, 
\nonumber\\\end{eqnarray}
with  
\begin{eqnarray}
\tilde\delta_\mp =  \frac{f_T^\perp}{f_T}  \frac{m_{q_1} \mp m_{q_2}}{m_T} \,, \quad 
\delta_\mp =  \frac{f_T}{f_T^\perp}  \frac{m_{q_1} \mp m_{q_2}}{m_T} \,.
\end{eqnarray}

Using the conformal basis, the leading-twist LCDAs $\phi_{\parallel,\perp}(u,\mu)$ can be expressed in a series of Gegenbauer polynomials. The LCDAs can be approximately expanded up to the term including the first Gegenbauer moment, $a_1^{\parallel,\perp}$, as
\begin{eqnarray}\label{eq:phi-as}
    \phi_{\parallel}(u) &=& 30 u(1-u)(2u-1) \frac{3}{5} a_1^\parallel, \\
    \phi_{\perp}(u) &=& 6u(1-u) + 30 u(1-u)(2u-1) \frac{3}{5} a_1^\perp, \label{eq:phi-as-2}
\end{eqnarray}
where the Gegenbauer moments renormalize multiplicatively:
  \begin{equation}
     \left( f^{(\perp)} a_\ell^{\parallel\,(\perp)} \right) (\mu) =
     \left( f^{(\perp)} a_\ell^{\parallel\,(\perp)} \right) (\mu_0)
  \left(\frac{\alpha_s(\mu_0)}{\alpha_s(\mu)}\right)^{-\gamma_{\ell}^{\parallel\,(\perp)}/{b}},
  \label{eq:RGparallel}
   \end{equation}
with $b=(11 N_c -2n_f)/3$ and the one-loop anomalous dimensions being \cite{Artru:1989zv,Gross:1973ju}
  \begin{eqnarray}
  \gamma_{\ell}^\parallel  = \frac{N_c^2-1}{2N_c}
  \left(1-\frac{2}{(\ell+1)(\ell+2)}+4 \sum_{j=2}^{\ell+1} \frac{1}{j}\right),
  \label{eq:1loopandim}
  \end{eqnarray}
\begin{eqnarray}
  \gamma_{\ell}^\perp  =\frac{N_c^2-1}{2N_c}
  \left(1+4 \sum_{j=2}^{\ell+1} \frac{1}{j}\right) \,.
  \label{eq:gamma_perp}
  \end{eqnarray}
For $\phi_{\parallel}$, we will lump  $3 a_1^\parallel/5$ into $f_T$, i.e., a new normalization with $a_1^\parallel=5/3$.  Note that here $a_1^\perp$ originates from the quark mass difference in the $2^{--}$ tensor meson, and give correction to $\phi_\perp$'s asymptotic form.

Taking the approximate leading-twist LCDAs, in the following sections, we will use the QCD sum rule approach to estimate the relevant parameters: the decay constants $f_T$ and $f_T^\perp$, and first Gegenbauer moment $a_1^\perp$.

\section{Calculations of $f_T, f_T^\perp$ and $a_1^\perp$}
\label{sec:calculation}
In this section, we calculate the parameters, $f_T$ and $f_T^\perp$, and $a_1^\perp$, that are relevant to determining the LCDAs of the $2^{--}$ tensor mesons, via the two-point correlation functions. First, we calculate the $f_T^\perp$ coupling constant. For this aim, we consider the following correlation function
\begin{equation}
   \label{eq:26}
   \Pi_{\mu \nu \delta \alpha \beta}^{\prime} = i \int d^4x \langle 0 | j_{\mu \nu \delta}^{\prime \dag} (x) j_{\alpha \beta} (y) |0 \rangle |_{y \rightarrow 0} e^{i qx } \,,
 \end{equation}
 where $j_{\alpha \beta}$ is the interpolating current,
\begin{equation}
  \label{eq:4}
  \begin{split}
    j_{\alpha \beta} & =  \big[ \bar{q}_1(y) \gamma_\alpha \gamma_5 \overleftrightarrow{\mathcal{D}_\beta} q_2(y) 
    + \bar{q}_1(y) \gamma_\beta \gamma_5 \overleftrightarrow{\mathcal{D_\alpha}} q_2(y) \big] \,,
\end{split}
\end{equation}
and
\begin{equation}
  \label{eq:20}
  j_{\mu \nu \delta}^{\prime \dag} = \bar{q}_2 (x)  \sigma_{\mu \nu} \gamma_5 i \widetilde{\overleftrightarrow{\mathcal{D}_\delta}} q_1(x)\,.
\end{equation}
The covariant derivative is defined as
\begin{equation}
  \label{eq:5}
  \begin{split}
    \overleftrightarrow{\mathcal{D_\alpha}} &=   \overrightarrow {\mathcal{D}_\alpha} - \overleftarrow{\mathcal{D}_\alpha}   \\
    &=   \overrightarrow{\partial_\alpha} - i \frac{g}{2} \lambda^a A^a_\mu - \overleftarrow{\partial_\alpha} - ig \frac{\lambda^a}{2} A_\mu^a  \,,  \\
    \widetilde{\overleftrightarrow{\mathcal{D}_\delta}} &= \overrightarrow{\partial}_\delta + \overleftarrow{\partial}_\delta \,,
     \end{split} 
\end{equation}
where $\lambda^a$ are the Gell-Mann matrices.
 The phenomenological part of the correlation function can be rewritten, by saturating it with the $2^{--}$ tensor mesons, as
\begin{equation}
  \label{eq:7}
    \Pi_{\mu \nu \delta \alpha \beta} = \frac{ \langle 0 | j_{\mu \nu \delta}^{\prime \dagger} | T \rangle \langle T | j_{\alpha \beta} |0 \rangle} {m_T^2 - q^2} + ... \,,
  \end{equation}
where $``..."$ denotes the higher states and continuum contributions, and the matrix elements are given by
  \begin{equation}
    \label{eq:8}
    \begin{split}
     \langle 0 | j_{\mu \nu \delta}^{\prime \dag} | T(q,\lambda) \rangle &=  f_T^\perp m_T^2 [\epsilon^{(\lambda)}_{\mu \delta} q_\nu - \epsilon^{(\lambda)}_{\nu \delta} q_\mu]  \,, \\
      \langle T(q, \lambda) | j_{\alpha \beta} |0 \rangle &= 2 f_T m_T^3 \epsilon^{(\lambda)*}_{\alpha \beta}  \,,
   \end{split}
 \end{equation}
 with $\epsilon^{(\lambda)}_{\alpha \beta}$ being the polarization tensor of the $2^{--}$ states.
Using the relation
 \begin{equation}
   \label{eq:9}
   \epsilon^{(\lambda)*}_{\mu \delta} \epsilon_{\alpha \beta} = \frac{1}{2} T_{\mu \alpha} T_{\beta \delta} + \frac{1}{2} T_{\mu \beta} T_{\delta \alpha} -\frac{1}{3} T_{\mu \delta} T_{\alpha \beta},
 \end{equation}
 where $T_{\mu \nu} = - g_{\mu \nu} + \frac{q_\mu q_\nu}{m_T^2}$, and considering only the coefficient of the Lorentz structure: $\Pi^{\prime}_{\mu \nu \delta \alpha \beta}  \to \Pi^{\prime} \big[(g_{\mu \alpha} g_{\delta \beta} + g_{\mu \beta} g_{\delta \alpha}) q_\nu - (\mu \leftrightarrow \nu) \big] $ for the phenomenological part given in Eq. (\ref{eq:7}), we get
 \begin{equation}
   \label{eq:32}
   \Pi^\prime = \frac{f_T f_T^\perp m_T^5}{(m_T^2 - q^2)}  + \dots   \,.
 \end{equation}
The choice of this structure is dictated by the fact that this structure contains only tensor meson contributions and free of spin-\rm{1} and spin-\rm{0} state contributions.

One should note that the $2^{-+}$ states can contribute this sum rule result by the G-parity violating effect, which is due to $m_{q_1}-m_{q_2}\neq 0$, since, due to the G-parity, in SU(3) limit, their $\phi_\parallel$ is symmetric under the replacement $u \to 1-u$, while $\phi_\perp$ is antisymmetric, in contrast to the case of $2^{--}$ states. Because the correction due to the lowest $2^{-+}$ state, entering this sum rule, is of order $(m_{q_1}-m_{q_2})^2$, we will neglect it.

To calculate the first Gegenbauer moment $a_1^\perp$ that gives corrections to asymptotic form of $\phi_\perp$, we consider the second correlation function introduced by
\begin{equation}
  \label{eq:3}
  \Pi_{\mu \nu \delta \alpha \beta} = i \int d^4x \langle 0 | j_{\mu \nu \delta}^\dagger (x) j_{\alpha \beta}(y) |0 \rangle |_{y \rightarrow 0} e^{iqx} \,,
\end{equation}
where $ j_{\mu \nu \delta}^\dag $ is given by
$
  j_{\mu \nu \delta}^\dag = \bar{q}_2(x) \sigma_{\mu \nu} \gamma_5 i \overleftrightarrow{\mathcal{D_\delta}} q_1 (x) \,,
$
and satisfies
  \begin{equation}
    \label{eq:8}
    \begin{split}
      \langle 0 | j_{\mu \nu \delta}^\dagger | T(q, \lambda) \rangle &=\frac{3}{5} a_1^\perp f_T^\perp m_T^2 (\epsilon^{(\lambda)}_{\mu \delta} q_\nu -\epsilon^{(\lambda)}_{\nu \delta} q_\mu)   \,.  \end{split}
 \end{equation}
 Using these definitions and separating the coefficient of the structure $ \big[ (g_{\mu \alpha} g_{\delta \beta} + g_{\mu \beta} g_{\delta \alpha}) q_\nu - (\mu \leftrightarrow \nu) \big]$,  we get the following expression for the correlation function in terms of hadronic degrees of freedom,
 \begin{equation}
   \label{eq:10}
   \begin{split}
     \Pi_{\mu \nu \delta \alpha \beta} = \frac{f_T^\perp m_T^2 f_T m_T^3 (\frac{3}{5} a_1^\perp)}{(m_T^2 - q^2)} \times 
     \bigg( (g_{\mu \alpha} g_{\delta \beta} + g_{\mu \beta} g_{\delta \alpha})q_\nu -(\mu \leftrightarrow \nu) + \text{Other structures} \bigg)  \,.
  \end{split}
 \end{equation}
We give a brief  introduction to the factor, $\frac{3}{5} a_1^\perp$ as follows. The matrix element of the non-local operator $\bar{u}(z+x) \sigma_{\mu \nu} \gamma_5 i \overleftrightarrow{\mathcal{D}_\alpha} d(-z+x)$ between the $2^{--}$ tensor meson and vacuum is defined as
 \begin{equation}
   \label{eq:19}
   \begin{split}
     &\langle T(P,\lambda) | \bar{q}_1 (z+x) \sigma_{\mu \nu} \gamma_5 i \overleftrightarrow{\mathcal{D}_\alpha} q_2 (-z+x) |0 \rangle  |_{x \rightarrow 0}
          =  f_{T}^\perp m_T^2 \big\{ \\
     &\hskip0.5cm  
     \big[ \epsilon^{(\lambda)*}_{\mu \beta} z^\beta P_\nu P_\alpha - \epsilon^{(\lambda)*}_{\nu \beta} z^\beta P_\mu P_\alpha \big] 
      \frac{1}{P z} \int_0^1 du e^{i (u - \bar{u}) P z} (2u -1) \phi_\perp(u) + ... \big\} |_{x \rightarrow 0}. 
 \end{split}
 \end{equation}
 After multiplying both sides of Eq.~(\ref{eq:19}) by $z_\alpha$, we get, with $x \to 0$,
 \begin{equation}
   \label{eq:21}
   \begin{split}
     \langle T(P,\lambda) | \bar{u} (z) \sigma_{\mu \nu} \gamma_5 i \overleftrightarrow{\mathcal{D}_\alpha} z^\alpha d(-z) |0 \rangle =
     -i f_T^\perp m_T \big[ \epsilon^{(\lambda)*}_{\mu \alpha} P_\nu - \epsilon^{(\lambda)*}_{\nu \alpha} P_\mu \big] z^\alpha \langle \xi \rangle \,,
   \end{split}
 \end{equation}
 where
 \begin{equation}
   \label{eq:22}
   \langle \xi \rangle = \int_0^1 du (2u - 1) \phi_\perp (u) \,.
 \end{equation}
The leading-twist LCDA $\phi_\perp (u)$ can be expanded in a series of the Gegenbauer moments as
 \begin{equation}
   \label{eq:23}
   \phi_\perp(u,\mu) = 6 u \bar{u} \sum_{l=0}^{\infty} a_l^\perp(\mu) C_l^{3/2} (2u-1) \,,
 \end{equation}
where  $C_l^{3/2} (2u-1)$ are the Gegenbauer polynomials, $\mu$ is the normalization scale, and the Gegenbauer moments are defined as
 \begin{equation}
   \label{eq:24}
   a_l^\perp(\mu) = \frac{2(2l +3)}{3(l+1)(l+2)} \int_0^1 du C_l^{3/2} (2u-1) \phi_\perp (u) \,.
 \end{equation}
From this expression, we get
 \begin{equation}
   \label{eq:25}
   a_1^\perp = \frac{5}{3} \langle \xi \rangle  \,,
 \end{equation}
 i.e. $\langle \xi \rangle = \frac{3}{5} a_1^\perp$.
 
Note that, for a $2^{--}$ state, $a_1^\perp$, contributing the G-parity violating correction to $\phi_\perp(u)$'s asymptotic LCDA, $6u(1-u)$, is due to the mass difference of the two quarks. However, for a $2^{--}$ state, although the corresponding coupling $f_T$ is G-violating, the current $ j_{\mu \nu \delta}^\dag $ couples to the $2^{-+}$ state in a G-parity conserving way.   As a result,  for the sum rule result derived from Eq.~(\ref{eq:3}), both contributions of the lowest $2^{--}$ and $2^{-+}$ states  are  the same order of magnitude, (${\cal O}(m_{q_1} -m_{q_2})$). Although the prediction of $a_1^\perp$ may have a large uncertainty, we will show later that the resulting $f_T \gg f_T^\perp$ implies that $\phi_\perp$ is negligible in the exclusive processes.

 The correlation functions can also be calculated in terms of quark-gluon fields using the operator product expansion (OPE). Using expressions of interpolating current $j_{\mu \nu \delta}$ and $j_{\alpha \beta}$ and contracting all quark pairs, we get following expression for the correlation function,
 \begin{equation}
   \label{eq:11}
   \begin{split}    
    \Pi_{\mu \nu \delta \alpha \beta}^\prime &= - i \int d^4x e^{iqx} \\
     & Tr \bigg[ S_{q_2}(y-x) \sigma_{\mu \nu} \gamma_5 i \widetilde{\overleftrightarrow{\mathcal{D}}_\delta}(x) S_{q_1}(x-y) \gamma_\alpha \gamma_5 \overleftrightarrow{\mathcal{D}_\beta}(y) + (\alpha \leftrightarrow \beta) \bigg] |_{y \rightarrow 0} \,,  \\     
     \Pi_{\mu \nu \delta \alpha \beta} &= - i \int d^4x e^{iqx} \\
     & Tr \bigg[ S_{q_2}(y-x) \sigma_{\mu \nu} \gamma_5 i \overleftrightarrow{\mathcal{D}_\delta}(x) S_{q_1}(x-y) \gamma_\alpha \gamma_5 \overleftrightarrow{\mathcal{D}_\beta}(y)   + (\alpha \leftrightarrow \beta) \bigg] |_{y \rightarrow 0}  \,. \\
   \end{split}
 \end{equation}
 The calculations are performed in the fixed-point gauge $x^\mu A^a_\mu =0$. In this gauge the gluon field is expressed via gluon field strength as
\begin{equation}
  \label{eq:6}
  A_\mu^a = \int dt x^\beta t G_{\beta \mu}^a (tx) = \frac{1}{2} x^\beta G_{\beta \mu}(0) + \frac{1}{3} x^\rho x^\beta \mathcal{D}_\rho G_{\beta \mu}^a(0) .
\end{equation}
 From this expressions it follows that for calculation of $\Pi_{\mu \nu \delta \alpha \beta}$ from the QCD side we need expression of the light quark propagator. For the light quark propagator  including operators up to dimension eight is given by (see for example~\cite{Lee:2002jb})
 \begin{equation}
   \label{eq:12}
   \begin{split}
     S_q^{ab}(z) &= \frac{i}{2 \pi^2 z^4} \slashed{z} \delta^{ab} - \frac{m_q}{2^2 \pi^2 z^2} \delta^{ab} - \frac{1}{12} \langle \bar{q} q \rangle \delta^{ab} + \frac{i}{2^4 3} m_q \langle \bar{q} q \rangle \slashed{z} \delta^{ab} \\
     & - \frac{z^2}{2^6 3} m_0^2 \langle \bar{q} q \rangle \delta^{ab} + \frac{i z^2}{2^7 3^2} m_q m_0^2 \langle \bar{q} q \rangle \slashed{z} \delta^{ab} - \frac{z^4}{2^{9} 3^3} \langle \bar{q} q \rangle \langle g^2 G^2 \rangle \delta^{ab} \\
     & + \frac{i}{2^5 \pi^2 z^2} (g_s G_{\alpha \beta})^n (\slashed{z} \sigma^{\alpha\beta} + \sigma^{\alpha \beta} \slashed{z}) \frac{\lambda_{ab}^n}{2} \\
     &- \frac{1}{2^5 \pi^2} m_q \bigg[ ln(-\frac{z^2 \Lambda^2}{4}) + 2\gamma_E \bigg] (g_s G_{\alpha \beta}^n) \frac{\lambda_{ab}^n}{2} \sigma^{\alpha \beta} ,
     % + \frac{1}{2^6 3} m_0^2 \langle \bar{q} q \rangle \sigma_{\alpha \beta} \frac{\lambda_{ab}^n}{2}\\
     % &- \frac{i}{2^8 3} m_q m_0^2 \langle \bar{q} q \rangle (\slashed{z} \sigma_{\alpha \beta} + \sigma_{\alpha \beta} \slashed{z}) \frac{\lambda_{ab}^n}{2} 
     % + \frac{z^2}{2^{10} 3^2} \langle g_s^2 G^2 \rangle \langle \bar{q} q \rangle \sigma_{\alpha \beta} \frac{\lambda_{ab}^n}{2} \,,
   \end{split}
 \end{equation}
 where $z = x-y$ and $\Lambda$ is the parameter separating perturbative and non-perturbative domains and we choose it $\Lambda = 0.5~\rm{GeV}$~\cite{Chetyrkin:2007vm}.  Compared with \cite{Lee:2002jb}, our coefficient of the dimension-7 term is larger by a factor 2.

 In calculations, we also need the expression for the vacuum expectation value (\rm{v.e.v}) of the quark and gluon field product. This \rm{v.e.v} is determined as
 \begin{equation}
   \label{eq:30}
   \begin{split}
   \langle 0 | T\{ q^a(x) g_s G_{\alpha \beta}^n \bar{q}^b(y) \}\rangle &= \frac{1}{2^6 3} m_0^2 \langle \bar{q} q \rangle \sigma_{\alpha \beta} \frac{\lambda_{ab}^n}{2} - \frac{i}{2^8 3} m_q m_0^2 \langle \bar{q} q \rangle (\slashed{z} \sigma_{\alpha \beta} + \sigma_{\alpha \beta} \slashed{z}) \frac{\lambda_{ab}^n}{2} \\
   &  + \frac{z^2}{2^{10} 3^2} \langle g_s^2 G^2 \rangle \langle \bar{q} q \rangle \sigma_{\alpha \beta} \frac{\lambda_{ab}^n}{2}   \,.
   \end{split}
 \end{equation}
Putting these expressions to the correlation functions and getting derivatives with respect to $x$ and $y$ and after performing derivatives over $y$ and letting $y=0$, we obtain the expression of the correlation function in the  coordinate representation. For obtaining expression of the correlation function in the momentum space, we perform integration over $x$ by using
 \begin{equation}
   \label{eq:13}
   \begin{split}
     I_0 &= \int d^4x \frac{e^{iqx}}{(x^2)^n} \\
     &=(-1)^n (-i) \frac{\pi^2}{\Gamma(n)} \int_0^{\infty} d\alpha~\alpha^{n-3} e^{-Q^2/4\alpha} \,,
   \end{split}
 \end{equation}
 where $Q^2 = -q^2$. The integrals $\int d^4x \frac{e^{iqx}}{(x^2)^n} x_\mu x_\nu x_\delta$ can be obtained from this integral by replacing $x_\mu \rightarrow -i \frac{\partial}{\partial q_\mu}$ and get from right side of $I_0$. Separating the coefficient of the structure $  \big( g_{\mu \alpha} g_{\beta \delta} q_\nu - g_{\nu \alpha} g_{\beta \delta} q_\mu \big)$ and performing the Borel transformation with the help of expression
 \begin{equation}
   \label{eq:14}
   \begin{split}
     \hat{B} e^{-Q^2/{4 \alpha}} &= \delta \Big(\frac{1}{M^2} - \frac{1}{4\alpha} \Big) \,, \\
     \hat{B} \frac{1}{m_T^2+Q^2} &= e^{-m_T^2/M^2} \,,
 \end{split}
\end{equation}
where $M^2$ is the Borel mass, we get the following two sum rules:
 \begin{subequations}
  \label{eq:15}
  \begin{align} \label{eq:15a}
    f_T^\perp f_T m_T^5 e^{-m_T^2/M^2} &= \frac{\langle g_s^2 G^2 \rangle}{192 \pi^2} \big( m_{q_1} +  m_{q_2} \big) \,,
\\
    \frac{3}{5} a_1^\perp f_T^\perp f_T m_T^5 e^{-m_T^2/M^2} &= \frac{1}{192 \pi^2 M^2} \bigg[ 2 M^2 \langle g_s^2 G^2 \rangle (m_{q_1}-m_{q_2}) \ln{\frac{\Lambda^2}{M^2}}  \nonumber \\
    &+\langle g_s^2 G^2 \rangle \big[ (1+ 6 \gamma_E) M^2 (m_{q_1}-m_{q_2}) - 24 \pi^2 \big( \langle \bar{q}_1 q_1 \rangle - \langle \bar{q}_2q_2 \rangle \big)\big] \nonumber \\
  &- 24 M^2 \big[ M^4 (m_{{q_1}} - m_{{q_2}}) - 4\pi^2 m_0^2 \big( \langle \bar{q}_1q_1 \rangle - \langle \bar{q}_2q_2 \rangle \big) \big]   \bigg] \,.   \label{eq:15b}
    \end{align}
 \end{subequations}
Here we have obtained the QCD OPE results at the quark-gluon level up to dimension 7. However, the sum rule for $f_T^\perp$ given in Eq.~(\ref{eq:15a}) has only one term proportional to the mass sum of the quark pair left, for which we have checked that both the OPE terms with dimension five (the quark-gluon mixing condensate) and dimension 7 (gluon condensate times the quark condensate) vanish, such that the subtraction of excited states also vanishes. In Eq.~(\ref{eq:15a}), the contributions arising from the $2^{-+}$ states are ${\cal O}(m_{q_1} -m_{q_2})^2$ can be reasonably negligible.  The $f_T^\perp$ result which vanishes in the chiral limit may imply the (transverse) production rate is small in $B$ decays.
For Eq.~(\ref{eq:15b}), its OPE result is proportional to $(m_{q_1} -m_{q_2})$ due to the G-parity violating effect, just as the previous discussion. 

After subtracting the contributions of higher states and continuum, i.e., by doing the replacement:
\begin{equation}
  \label{eq:16}
  (M^2)^n \rightarrow \frac{1}{\Gamma(n)} \int_0^{s_0} e^{-s/M^2} s^{n-1} ds  \,,
\end{equation}
one can easily obtain two sum rule results for the determination of $f_T^\perp$ and $a_1^\perp$ from Eqs.~(\ref{eq:15a}) and (\ref{eq:15b}).

For determination of $f_T^\perp$ and $a_1^\perp$, we need to know $f_T$. This coupling  $f_T$ can be obtained from the following two-point function,

\begin{equation}
  \label{eq:17}
\Pi_{\mu \nu \alpha \beta} = i \int d^4x  e^{i(qx)} \langle 0 | T \{j_{\mu \nu}(x) j_{\alpha \beta}^\dagger (y) \} |0 \rangle |_{y\to 0} \,.
\end{equation}
At the hadronic level,  after Borel transformation, we get 
\begin{equation}
  \label{eq:18}
  \Pi_{\mu \nu \alpha \beta} = 2 f_T^2 m_T^6 e^{-m_T^2/M^2} \bigg[ (g_{\mu \alpha} g_{\nu \beta} + g_{\mu \beta} g_{\nu \alpha} ) + \text{other structures} \bigg].
\end{equation}
Taking into account the coefficient of the structure $(g_{\mu \alpha} g_{\nu \beta} + g_{\mu \beta} g_{\nu \alpha})$, and the subtracting the contribution from higher states from the quark-gluon level with the replacement given in Eq.~(\ref{eq:15}), we get the desired sum rule for $f_T$:
\begin{equation}
  \label{eq:28}
  2 f_T^2 m_T^6 e^{-m_T^2/M^2} = \Pi^B.
\end{equation}
The invariant function $\Pi^{B}$ was calculated in~\cite{Chen:2011qu}, and, for this reason, we do not present its explicit expressions.
Just as Eq.~(\ref{eq:26}), we neglect the correction due to the lowest $2^{-+}$ state,  because it is of order $(m_{q_1}-m_{q_2})^2$.

\section{Numerical Analysis}
\label{sec:numerical}
In this section, we present our numerical analysis of the sum rules for the decay constants $f_T$, $f_T^\perp$, and $a_1^\perp$. The main input parameters entering to the sum rules for $f_T$, $f_T^\perp$, and $a_1^\perp$ are the quark masses and various condensates. Their values that we use are given in Table~\ref{tab:constant_values}~\cite{Olive:2016xmw,Ioffe:2005ym}.
\begin{table}[t]
  \centering
%  \begin{center}
\begin{tabular}{cc}
  \toprule
  % \rowcolor{yellow}
 % \multicolumn{2}{c}{Numerical Values} \\ \cmidrule{1-2}
  $m_q(\rm{2~GeV}) =  \frac{m_u + m_d}{2} = 3.5^{+0.7}_{-0.3}\times 10^{-3}~\rm{GeV}$ &  \\
% $m_u(\rm{2~GeV}) =  (2.9 \pm 0.6) \times 10^{-3}~\rm{GeV}$ &  \\
% $m_d(\rm{2~GeV}) =  (5.2 \pm 0.9) \times 10^{-3}~\rm{GeV}$ & \\
  $m_s(\rm{2~GeV}) =  (96^{+8}_{-4}) \times 10^{-3}~\rm{GeV}$ & \\
  $\langle \bar{q}q \rangle = -(0.23 \pm 0.03)^3~\rm{GeV^3}$ & \\
  $m_0^2 = (0.8 \pm 0.2)~\rm{GeV^2}$ &\\
  $\langle \bar{s} s \rangle = (0.8 \pm 0.2 )\langle \bar{q}q \rangle$ \\
  $\langle \frac{\alpha_s}{\pi} G^2 \rangle = (0.0012 \pm 0.0004)~\rm{GeV^4}$\\
  \bottomrule
\end{tabular}
%\end{center}
\caption{The values of input parameters.}
\label{tab:constant_values}
\end{table}
Here the masses of quarks are presented in $\overline{\text{MS}}$ scheme.  For simplicity, neglect the possible mixtures between two strange states, $K_2 (1820)$ (the $^1D_2$ state) and $K_2(1770)$ (the $^3 D_2$ state), and between ${\bar s}s $ and ${\bar u} u + {\bar d} d$ states.
  We use the $2^{--}$ meson masses with errors obtained in~\cite{Chen:2011qu}  as inputs. Assuming that the widths of these states are about $\lesssim$ 300 MeV, which is compatible with that of $K_2(1820)$ and other higher resonance states. To take into account the width uncertainty, in the QCD sum rule analysis we thus add the uncertainty $\pm 150$~MeV to the excited state threshold $\sqrt{s_0}$ that models the contribution from higher states as given in Eq.~(\ref{eq:16}).

\begin{figure}[h]
  \centering
  \includegraphics[scale=0.60]{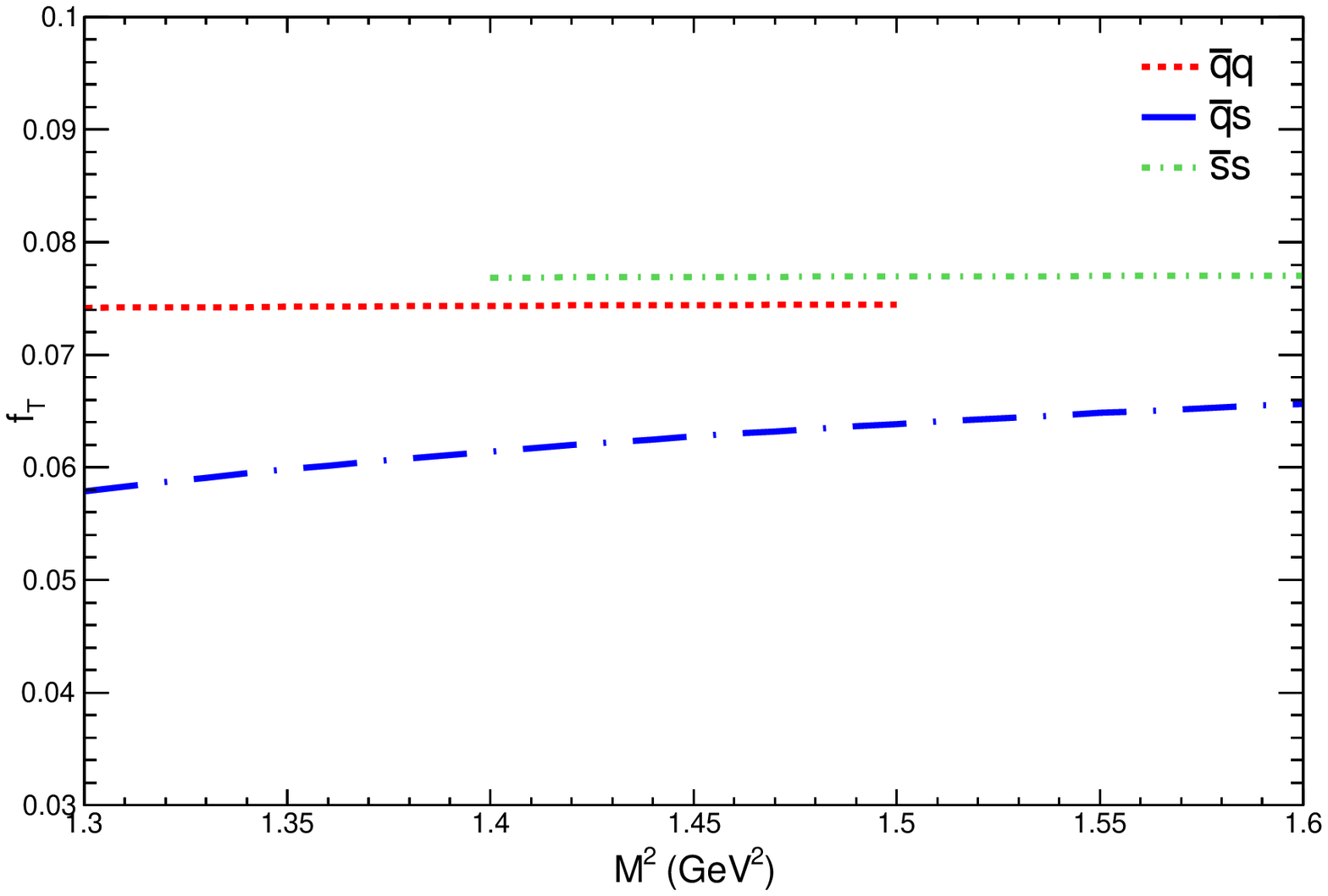}
  \caption{The dependence of $f_T$ on $M^2$.}
  \label{fig:fig1}
\end{figure}

In the all sum rule analyses,  the window of the Borel mass parameter and the central values of the continuum threshold $s_0$ that we adopt are shown in Table~\ref{tab:table2}, that were obtained from the analysis of the relevant mass sum rules presented in~\cite{Chen:2011qu}.
%%%%
\begin{table}[t]
  \centering
%  \begin{center}
\begin{tabular}{cccc}
  \toprule
  % \rowcolor{yellow}
 % \multicolumn{3}{c}{Numerical Values} \\  \cmidrule{1-3}
&\text{\hskip1cm}  $s_0~(\rm{GeV^2})$\text{\hskip1cm}  & Window $M^2~(\rm{GeV^2})$\text{\hskip1cm}   & $m_T~(\rm{GeV})$\\
  \midrule
  $\bar{q}q$ & $2.1^2$ & $1.3 - 1.6$ & $1.78 \pm 0.12$ \\
  $\bar{q}s$ & $2.2^2$ & $1.3 - 1.7$ & $1.85 \pm 0.14$ \\
  $\bar{s}s$ & $2.4^2$ & $1.4 - 1.8$ & $2.00 \pm 0.16$\\
  \bottomrule
\end{tabular}
%\end{center}
\caption{The values of $s_0$, ranges of the Borel mass squared $M^2$, and mass for different tensor mesons. These values are taken from~\cite{Chen:2011qu}.}
\label{tab:table2}
\end{table}
%%%%
 The dependence of $f_T$ on $M^2$ is illustrated in Figure~\ref{fig:fig1} at the central values of $s_0$.  From Figure~\ref{fig:fig1}, we show that the coupling constant $f_T$ exhibits good stability with respect to variation of $M^2$ in the working region. We get the following values:
\begin{equation}
  f_T = \begin{cases}
    (7.7 \pm 0.1) \times 10^{-2}, & \text{for $\bar{s}s$ } ,\\
    (6.2 \pm 0.4) \times 10^{-2}, & \text{for $\bar{q}s$ }  , \\
   (7.4 \pm 0.1) \times 10^{-2}, & \text{for $\bar{q}q$ }.
  \end{cases}
\end{equation}
Here and for the results in the following calculations, the errors correspond to the uncertainties due to variation of input parameters, threshold for higher states $s_0$,  theoretically predicted values for masses of tensor mesons, and Borel mass. 

\begin{figure}[!h]
  \centering
  \includegraphics[scale=0.60]{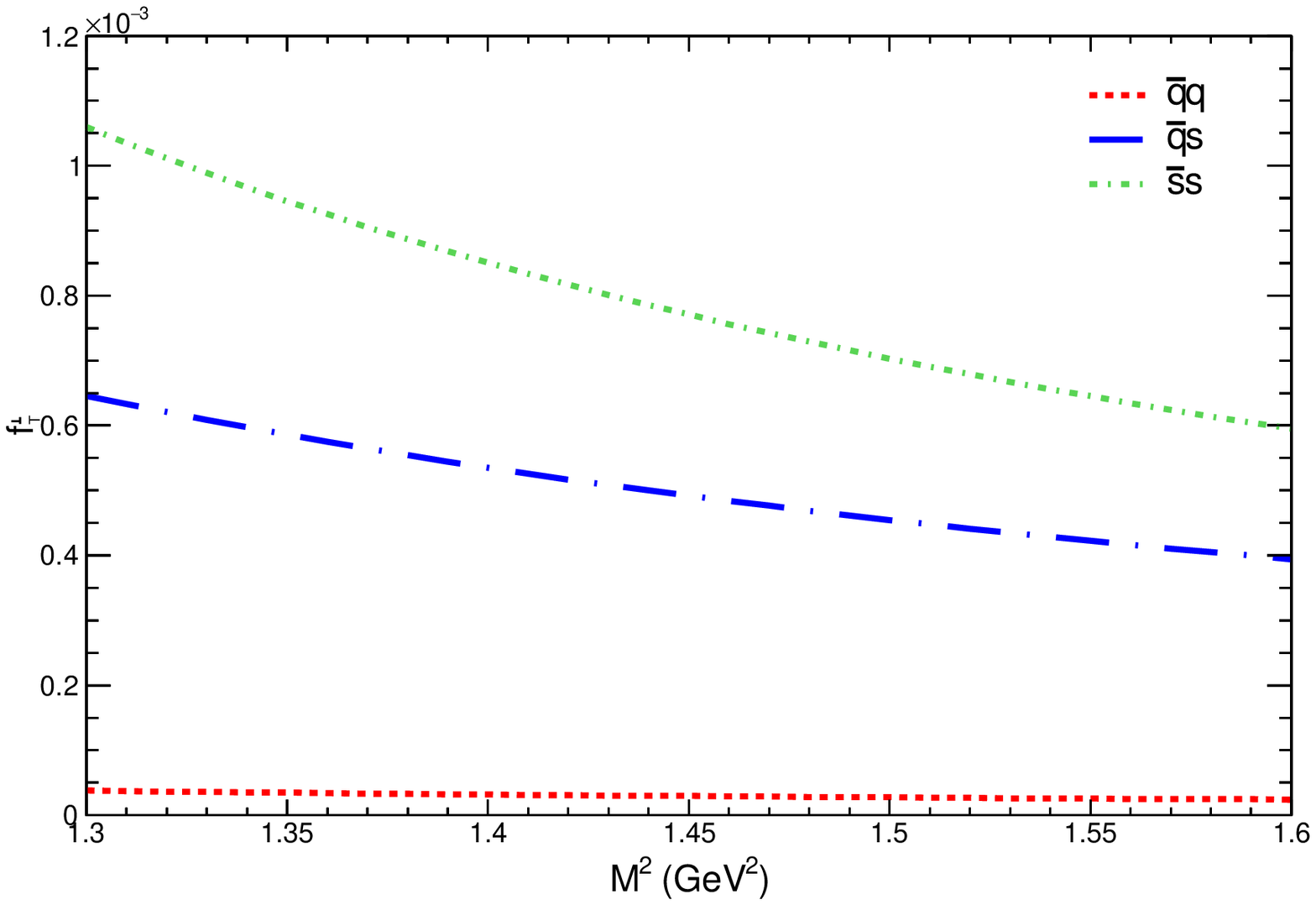}
  \caption{Same as in Figure~\ref{fig:fig1}, but for $f_T^\perp$.}
  \label{fig:fig2}
\end{figure}

Using the obtained values of $f_T$,  we further estimate $f_T^\perp$.  In Fig~(\ref{fig:fig2}), we show the dependence of $f_T^\perp$ for the tensor mesons with quark content $\bar{q}q$, $\bar{q}s$ and $\bar{s}s$ on $M^2$, respectively. 
 For the chosen Borel windows, we find that the contribution arising from higher states and continuum is less than $30\%$ of total results, and the non-perturbative contributions are less than $40\%$ of the perturbative one in OPE for the sum rule result given in Eq.~(\ref{eq:15b}). The resulting $f_T^\perp$ exhibits very good stability with respect to the variation of $M^2$. We find that
\begin{equation}
  f_T^\perp = \begin{cases}
    (8.2 \pm 2.2) \times 10^{-4}, & \text{for $\bar{s}s$ } ,\\
    (5.2 \pm 1.2) \times 10^{-4}, & \text{for $\bar{q}s$} \,,\\
    (3.5 \pm 1.1) \times 10^{-5}, & \text{for $\bar{q} q$} \,.
  \end{cases}
\end{equation}
Our results show that $f_T^\perp$ is much smaller than the corresponding $f_T$. The reason is that although we have performed the QCD OPE calculation at the quark-gluon level up to dimension 7, nevertheless, the sum rule for $f_T^\perp$ given in Eq.~(\ref{eq:15a}) has only one nonzero term left, which is proportional to the mass sum of the quark pair.  Phenomenologically, the smallness of $f_T^\perp$ result, which vanishes in the chiral limit, may imply that the (transverse) production rate is small in $B$ decays.

\begin{figure}[!h]
  \centering
  \includegraphics[scale=0.60]{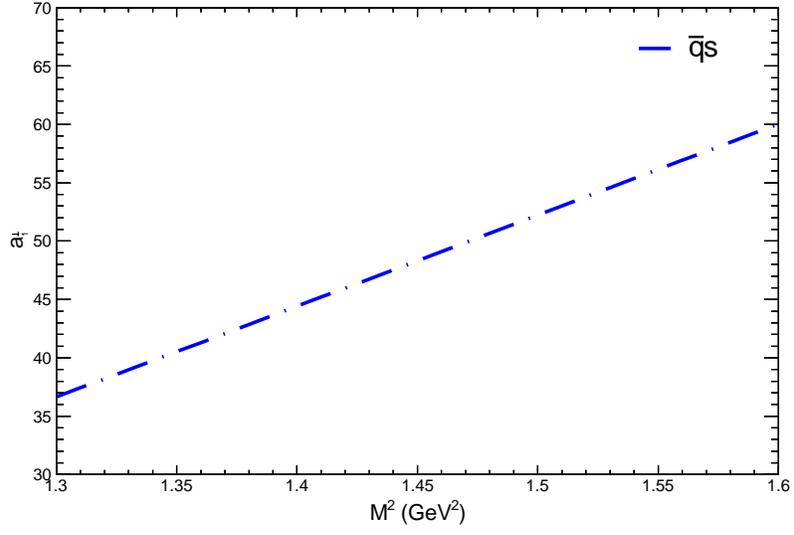}
  \caption{Same as in Figure~\ref{fig:fig1}, but for $a_1^\perp$.}
  \label{fig:fig3}
\end{figure}

The correction to the asymptotic form, $6u(1-u)$, of  the LCDAs $\phi_\perp(u)$ in a series of Gegenbauer polynomials is due to the mass difference of the two quarks in a $2^{--}$ state, and is given by $18 a_1^\perp u(1-u)(2u-1)$, where $a_1^\perp$  is the so-called first Gegenbauer moment. Considering only corrections due to $m_s$,  we illustrate the dependence of $a_1^\perp$ on $M^2$ using the central values of $s_0$ in Figure~\ref{fig:fig3}, and get the result,
\begin{equation}
  \label{eq:29}
  a_1^\perp =
    (48 \pm 12), \  \  \text{for $\bar{q}s$}\,,
\end{equation}
to which the $2^{-+}$ state also contribute the same order of magnitude;  thus, the value of $a_1^\perp$ for the lowest $2^{--}$ state may be overestimated by a factor $1\sim 2$.  
Because the resulting $(f_T^\perp / f_T)^2 \ll1$ and $(\frac{3}{5} a_1^\perp f_T^\perp / f_T)^2 \ll 1$ imply that $\phi_\perp$ is negligible in the exclusive processes.

\section{Summary}\label{sec:summary}

The  LCDAs of the light $2^{--}$ meson state are relevant ingredients to theoretically predict  $B$ decay channels involving the meson in B factories. 
The observation of the the light $2^{--}$ meson state is important from the viewpoint of the QCD-based quark model, and can help to clarify the rate deficit and polarization puzzles in $B$ decays involving vector or tensor meson(s) in the final states.

We have defined two-quark light-cone distribution amplitudes for the $1^3D_2$ light tensor meson states with quantum number $J^{PC}=2^{--}$.   In terms of twist-2 ones through the Wandzura-Wilczek relations, we have shown the results of the twist-3  two-quark LCDAs with quark mass corrections.  In the SU(3) limit, because of the G-parity, the chiral-even two-quark light-cone distribution amplitudes of this tensor meson are antisymmetric under the interchange of momentum fractions of the quark and antiquark, while the chiral-odd ones are symmetric.

Using QCD conformal partial expansion, we have shown the asymptotic leading twist  LCDAs with the strange quark mass correction and up to the term containing the first Gegenbauer moment (see Eqs.~(\ref{eq:phi-as}) and (\ref{eq:phi-as-2})). The relevant parameters: the decay constants $f_T$ and $f_T^\perp$, and first Gegenbauer moment $a_1^\perp$, are estimated by means of the QCD sum rule method.  The smallness of $f_T^\perp$ result may phenomenologically imply that the transverse production rate is small in $B$ decays. 
These parameters play a central role in the investigation of $B$ meson decaying into these negative parity tensor mesons which we are planning to study in the near future.

\section*{Acknowledgments}
K.C.Y.  is supported in part by the Ministry of Science and Technology of the Republic of China under Grant No. 102-2112-M-033-007-MY3.
\clearpage
\section*{References}

\bibliography{mybibfile}
\clearpage


\clearpage

\end{document}